\documentclass{PHYEAUTH}
\usepackage{graphicx}
\usepackage{amsmath}
\usepackage{amssymb}

\begin{document}

\begin{frontmatter}

\title{Quantum Monte Carlo Simulation of Exciton-Exciton\\
 Scattering in a GaAs/AlGaAs Quantum Well }
\date{29 July 2005}
\author{J. Shumway\thanksref{thank1}}
\address{Dept. of Physics and Astronomy, Arizona State University, Tempe, AZ 85287-1504}
\thanks[thank1]{Email: shumway@mailaps.org}

\begin{abstract}
We present a computer simulation of exciton-exciton scattering in
a quantum well. Specifically, we use quantum
Monte Carlo techniques to study the bound and continuum states of two excitons
in a 10 nm wide GaAs/Al$_{0.3}$Ga$_{0.7}$As quantum well. From these 
bound and continuum states we
extract the momentum-dependent phase shifts for s-wave scattering.
A surprising finding of this work is that a commonly studied effective-mass model for
excitons in a 10 nm quantum well actually supports {\em two} bound biexciton
states. The second, weakly bound state may dramatically enhance 
exciton-exciton interactions. We also fit our results to a hard-disk model
and indicate directions for future work.
\end{abstract}

\begin{keyword}
Excitons\sep Quantum wells \sep Quantum Monte Carlo  \sep Scattering
\PACS 78.67.De \sep 02.70.Ss \sep 71.35.Ay 
%  78.67: Optical properties of low-dimensional, mesoscopic, and nanoscale 
% materials and structures - Quantum wells
% 02.70.Ss: Quantum Monte Carlo methods 
% 71.35.Ay: Excitons and related phenomena
\end{keyword}
\end{frontmatter}

%[main text]

\section{Introduction}
Exciton-exciton interactions in quantum wells are becoming increasingly
important to science and technology. For example, the interaction of two
excitons is a mechanism for non-linear optical response
\cite{Hiroshima:1988}. It is the excitonic
component of polaritons that lead to their interactions. Also, in experimental
studies of cold excitonic gases in quantum wells, exciton-exciton scattering
is a process that drives the excitons towards a Bose distribution.

Exciton-exciton interactions are difficult to theoretically predict \cite{Shumway:2001b}.
Fast collisions between excitons may be treated with time dependent
perturbation theory \cite{Cuiti:1998}. Carrier exchange is known to be a significant contribution
to the scattering process \cite{Cuiti:1998}. At a different extreme, slow moving excitons interact
through dipole interactions. If the excitons have intrinsic dipoles, perhaps
from carefully engineered quantum well bandstructure or an applied electric
field, the repulsive dipole-dipole force can dominate interactions
\cite{Hiroshima:1988,Koh:1997}. In the more
common case of unpolarized excitons, quantum fluctuations of the polarization
leads to attractive van der Waals interactions, which are often neglected.
Finally, we know that the wavefunction describing elastic exciton collisions
must be orthogonal to any bound biexciton states. The presence of weakly
bound biexciton states can have dramatic effect on exciton scattering cross-sections
\cite{Shumway:2001b}.

In this work we have extended a quantum Monte Carlo technique in order to 
to study exciton scattering in a GaAs/Al$_{0.3}$Ga$_{0.7}$As quantum wells.
An earlier paper studied bulk exciton-exciton scattering for a generic
model of a semiconductor \cite{Shumway:2001b}. The simulation technique directly
samples the energy of the four-particle scattering wavefunction.
This allows all mechanisms listed in the previous paragraph
--- carrier exchange, dipole and van der Waals forces, and orthogonality
to bound biexcitons --- to be fully incorporated in the numerical simulation.
Random walks are used to project essentially exact energies from
a set of trial wavefunctions, thus providing numerical estimates of
the phase shifts and scattering cross sections for elastic exciton-exciton
collisions. Our new calculations apply this technique to a quantum
well, allowing a detailed study of the exciton-exciton collision process
within the effective mass approximation. Two key results of these calculations
are the observation of a second, weakly-bound biexciton and the numerical
estimate of a large cross-section $\sigma \gtrsim 100$ nm for low-energy 
exciton-exciton scattering in a quantum well.

In the next section of this paper, we describe the effective mass model we use. After that
we give a brief description of our computational technique and details of our
simulation of a GaAs/Al$_{0.3}$Ga$_{0.7}$As well. The last section is a discussion 
that emphasizes new results of this paper, some subtleties of two-dimensional scattering,
and future directions for this research.

\section{Model}

We use a single-band effective mass model,
\begin{equation}
\begin{split}
H= &-\frac{\hbar^2}{2m_e^*}(\nabla_1^2 + \nabla_2^2)
 -\frac{\hbar^2}{2m_h^*}(\nabla_a^2 + \nabla_b^2)\\
 &+\frac{e^2}{\epsilon}\left(
 \frac{1}{r_{12}}+\frac{1}{r_{ab}}
 -\frac{1}{r_{1a}}-\frac{1}{r_{2b}}
 -\frac{1}{r_{1b}}-\frac{1}{r_{2a}}\right)\\
 &+V_e(z_1)+V_e(z_2)+V_h(z_a)+V_h(z_b)
\end{split}
\end{equation}
Here the electrons are labeled $1$ and $2$ and the holes are labeled $a$ and $b$.
The isotropic effective masses are $m_e^*=0.0667 m_0$ and $m_h^*=0.34 m_0$,
and the dielectric constant is $\epsilon=12$. (Future calculations will include
the anisotropy of the heavy-hole mass.)  The confining potentials $V_e$ and
$V_h$ are finite square wells with offsets $\Delta V_e = 216$ eV and
$\Delta V_h  = 133$ meV. These material constants have been chosen
to match other theoretical studies of excitons and biexcitons in GaAs/Al$_{0.3}$Ga$_{0.7}$As
quantum wells for direct comparison with this work \cite{Riva:2000,Filinov:2003b}.

The electron effective mass and dielectric constant define natural
length and energy scales for the problem: the donor 
effective Bohr radius, $a_B^* = \hbar^2\epsilon/m_e^*e^2=9.97$ nm,
and the donor effective Rydberg, $\text{Ryd}^*=\hbar/2m_ea_B^*=5.79$ meV.
In this paper we present numerical answers to three digits for comparison to
other theoretical papers; such accuracy is beyond the limits of the effective mass model
and the leading digit should suffice for comparison to experiments.

For a well of width 9.97 nm, we find an exciton binding energy of 9.22 meV,
using QMC methods described in the next section.
Previously published QMC calculations report energies of about 10.0 meV for 
the exciton binding energy at this well width \cite{Riva:2000,Filinov:2003b}.
We have also calculated the
biexciton binding energy, which is around 1.87$\pm$0.01 meV, in very good agreement with
experimental \cite{Adachi:1997} and theoretical \cite{Tsuchiya:2004} values.

We also see a second, very weakly-bound excited biexciton state with binding energy 0.032$\pm$0.007 meV. This state is not likely to be directly observable, but it can have a big influence
on the scattering length. Also, note that this calculation is within the effective-mass approximation
with isotropic effective masses, and the bound state may be an artifact of this approximation.
Nevertheless, this results suggests that a second bound biexciton state is possible
in a GaAs/AlGaAs quantum well.

\begin{figure}[t]
\begin{center}\leavevmode
\includegraphics[width=\linewidth]{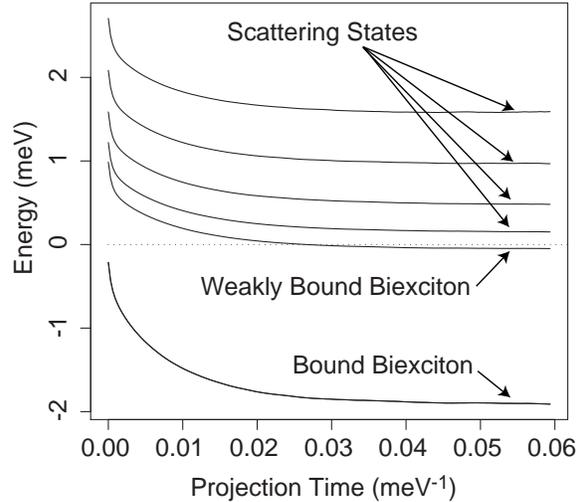}
\caption{Projection of eigenvalues with a random walk in imaginary
time (shown for the  simulation with $r_{\text{cut}}=180$ nm). 
The projection over an imaginary time 0.06 mev$^{-1}\cdot\hbar$ takes
a buffer of 200 Monte Carlo steps.}
\label{fig:proj}\end{center}\end{figure}

\section{Calculation Technique and Results}
We use random walks to project a set of wavefunctions in imaginary time, as described 
in Ref.~\cite{Shumway:2001b}. We first construct variational wavefunctions corresponding 
to a bound state and excited state of scattering. The excitons can be paired two ways:
electron $1$ can be paired with hole $a$ and electron $1$ with hole $b$, which we
denote $1a,2b$, or we can have the pairing $1b,2a$.
Eigenstates are symmetric or antisymmetric combinations,
\begin{equation}
\Phi_{\pm\pm} = \Phi_{1a,2b} \pm \Phi_{1b,2a}.
\end{equation}
These wavefunctions can be related to spin states through the symmetry of the wavefunction.
In these calculations we are just studying the $\Phi_{++}$ wavefunction, since this is the one that has the bound biexciton. This is one channel for scattering, and to get cross sections for different exciton spin states we will need to study $\Phi_{--}$ also.

Next we discretize the scattering states by restricting the excitons to never be separated by a distance greater than $R_n$. We choose values of $R_n$ between 160 nm and 280 nm. This turns the scattering states into standing waves we can study with QMC.
Using QMC, we project the wavefunction with a random walk. The random walk is repeated projection of the wavefunction with $\exp{-\tau H}$. In Fig. \ref{fig:proj} 
we show the projection of eigenvalues from the simulation with $R_n$=200 nm.
We clearly see the bound biexciton projected to an energy of about -1.9 meV. 
There is also a much more subtle feature: a second weakly bound biexciton state. 
This can be thought of as a quantized vibrational excitation in the radial
exciton-exciton separation coordinate.

\begin{figure}[t]
\begin{center}\leavevmode
\includegraphics[width=\linewidth]{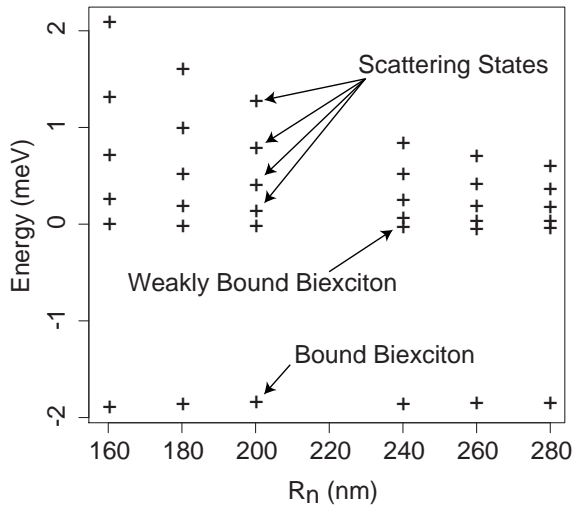}
\caption{Energy vs. cutoff. Small, random fluctuations are the statistical
noise in the Monte Carlo simulations.}
\label{fig:evsr}\end{center}\end{figure}

\begin{figure}[t]
\begin{center}\leavevmode
\includegraphics[width=\linewidth]{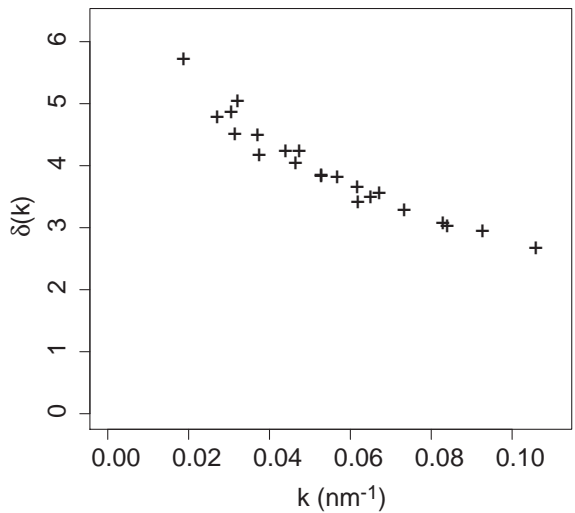}
\caption{Phase shifts calculated from data points in Fig.~\ref{fig:evsr}.}
\label{fig:deltak}\end{center}\end{figure}

In Fig. \ref{fig:evsr} we show the energy eigenvalues for a series of different $R_n$ values. 
We see that the two bound states are relatively insensitive to the boundary condition, while 
the scattering states systematically decrease in energy with increasing $R_n$.
These energy curves hold all the information about low energy s-wave scattering.
Since we know the energies of the scattering states, we know the relative momentum k. 
Because we know the scattering function has a node at $R_n$, we can infer the phase shift, 
$\delta(k)$. We plot the phase shift in Fig.~\ref{fig:deltak}. Again, the weakly bound vibrationally excited 
biexciton shows up, this time as the phase tends to $2\pi$, indicating two bound states.
The scatter in this data is the Monte 
Carlo noise from our simulations. 
\begin{figure}[t]
\begin{center}\leavevmode
\includegraphics[width=\linewidth]{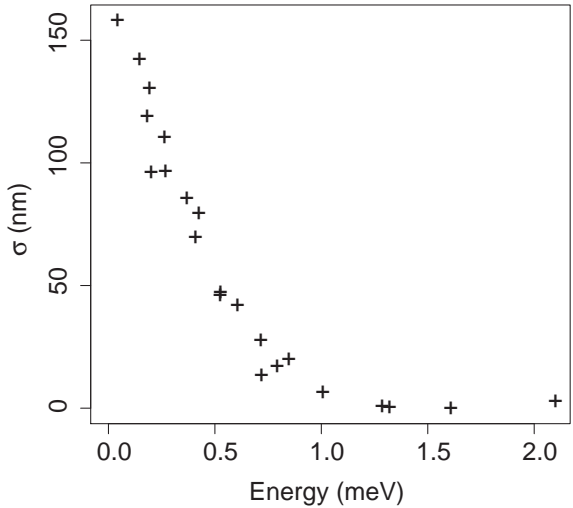}
\caption{Two-dimesional scattering cross sections
(with dimension of length) from data points in Fig.~\ref{fig:deltak}.}
\label{fig:sigmae}\end{center}\end{figure}

From the phase shift, we calculate the s-wave scattering cross sections, given by 
$\sigma=4\sin^2(\delta)/k$.
Fig.~\ref{fig:sigmae} shows the scattering cross sections from our raw Monte Carlo data.
At low energy, the cross section becomes quite large, $\sigma\gtrsim 100$ nm. Two
effects may contribute: (1) the weakly-bound excited biexciton state at $\sim 0.03$ meV
can enhance scattering, and (2) two dimensional systems have a logarithmic divergence 
of the scattering cross-section at
low energy. The statistical error in these simulations prevents us from distinguishing
between these contributions.

\section{Discussion}

Scattering in two dimensions differs significantly from three-dimensional
scattering \cite{Adhikari:1986b}. In particular, the scattering length
can no longer be simply obtained from an effective range expansion \cite{Adhikari:1986a}.
In three dimensions, the scattering length characterizes the interaction strength
for low energy collisions. The corresponding quantity in two dimensions is dimensionless:
the scattering is characterized by an effective range and a dimensionless parameter.
As a simple analysis, we will extract just one parameter from our results by fitting
our phase shifts to a hard-disk model. This analysis cannot capture resonant effects from the
weakly bound state, but it does provide a useful length scale.

The phase shift for s-wave scattering for hard disks of radius $a$ in two dimensions is
\begin{equation}\label{eq:disk}
\frac{\pi}{2}\cot\delta = \ln\left(\frac{ka}{2}\right)+\gamma+\frac{1}{4}a^2k^2 + \mathcal{O}(k^4),
\end{equation}
where $\gamma=0.5772\ldots$ is the Euler-Mascheroni constant.
A crude fit of our calculated values of $\frac{\pi}{2}\cot\delta$ to the hard disk 
formula gives $d\approx 2.8 a_B^*$ with a statistical uncertainty of about 10\%.
This suggests the excitons can be approximated as hard disks with radii of 27 nm,
if the electron and hole spins are each in singlets as required by our $\Phi_{++}$
wavefunction.

Now we must consider spin. We have only simulated the $\Phi_{++}$ scattering states.
In bulk, three-dimensional calculations, the $\Phi_{--}$ have a much smaller contribution
to the scattering, since there is now bound biexcition when electrons and holes are
spin polarized \cite{Shumway:2001b}. Using Table III of Ref.~\cite{Shumway:2001b}
and neglecting the $\Phi_{--}$ contribution, we
approximate the scattering cross section of paraexcitons as $\frac{1}{16}$ the symmetric
value, or 1.7 nm. Two orthoexictons in a relative $S=0$ state have a scattering
cross section that is  $\frac{1}{2}$ the symmetric value, or 13.5 nm. Recall that this
hard disk analysis neglects resonant effects from the weakly bound excited biexciton.

Future work will include a study of $\Phi_{++}$ and $\Phi_{--}$ scattering channels
for different geometries. We will also attempt to extract two parameters, 
an effective range and a scattering strength, from our calculations. Such
a parameterization could provide input for models of thermal equilibration 
rates and properties a weakly-interacting Bose condensate of excitons.
%\bibliographystyle{elsart-num}
%\bibliography{dots}

\end{document}